\RequirePackage{etoolbox}
\patchcmd{\bibliographystyle}{#1}{abbrvnat}{}{}
\documentclass[hf]{ceurart}
\pdfoutput=1
\usepackage[utf8]{inputenc}
\usepackage[T1]{fontenc}
\usepackage{cwdefs}
\usepackage{caption}
\usepackage{amssymb}
\usepackage{amsfonts}
\usepackage{amsmath}
\usepackage{bm}
\usepackage{xspace}
\usepackage{booktabs}
\usepackage{tikz}
\usetikzlibrary{matrix,snakes,arrows,shapes,positioning,arrows.meta,shapes.geometric}

\hypersetup{breaklinks=true}
\newcommand{\mgu}{\f{mgu}}

\newcommand\subsumedBy{\mathrel{\ooalign{$\geq$\cr
      \hidewidth\raise.225ex\hbox{$\cdot\mkern7.0mu$}\cr}}}
\newcommand\subsumes{\mathrel{\ooalign{$\leq$\cr
      \hidewidth\raise.225ex\hbox{$\cdot\mkern2.0mu$}\cr}}}
\newcommand\strictlySubsumedBy{\mathrel{\ooalign{$\geq$\cr
      \hidewidth\raise.225ex\hbox{$\cdot\mkern7.0mu$}\cr}}}
\newcommand\variant{\mathrel{\ooalign{$=$\cr
      \hidewidth\raise.7ex\hbox{$\cdot\mkern4.5mu$}\cr}}}

\newcommand{\OTTER}{\name{OTTER}\xspace}

\newcommand{\CMProver}{\name{CMProver}\xspace}
\newcommand{\SETHEO}{\name{SETHEO}\xspace}
\newcommand{\leanCoP}{\name{leanCoP}\xspace}
\newcommand{\PIE}{\name{PIE}\xspace}
\newcommand{\CDTOOLS}{\name{CD~Tools}\xspace}
\newcommand{\SGCD}{\name{SGCD}\xspace}

\newcommand{\TPTP}{\name{TPTP}\xspace}
\newcommand{\TPTPCDT}{\mbox{\name{TPTPCD2}}\xspace}
\newcommand{\CCS}{\mbox{\name{CCS}}\xspace}

\newcommand{\D}{\f{D}}

\renewcommand{\c}[1]{\mbox{\textsf{\textbf{#1}}}}
\renewcommand{\c}[1]{\bm{\mathsf{#1}}}
\renewcommand{\rimp}{\rightarrow_{\mathrm{rew}}}
\newcommand{\ldot}{{\,.\,}}
\newcommand{\oimp}{\imp}

\newcommand{\nimp}{{\rightarrow}}

\newcommand{\at}{{:}}
\newcommand{\derives}{\vdash}
\renewcommand{\P}{\f{P}}

\newcommand{\tl}{1.7cm}

\newcommand{\XC}{\textit{XC}}
\newcommand{\GS}{\textit{GS}}
\newcommand{\LC}{\textit{LC}}
\newcommand{\MC}{\textit{MC}}
\newcommand{\SC}{\textit{SC}}

\newcommand{\PMC}{\textit{M}\xspace}

\newcommand{\begindg}{\begin{tikzpicture}[>={Stealth[scale=1.2]},line join=bevel,]}
\newcommand{\rnode}[1]{\rule{0pt}{10pt}#1}
\makeatletter
\RenewDocumentCommand \maketitle {} {
  \thispagestyle{plain}
  \xdef\firstpage{\thepage}
  \ifbool{longmktitle}
  {
    \LongMaketitleBox
    \ProcessLongTitleBox
  }
  {
    \ifbool{dc}
    { \twocolumn[\MaketitleBox] }
    { \MaketitleBox }
    \printFirstPageNotes
  }
  \normalcolor \normalfont
  \renewcommand\thefootnote{\arabic{footnote}}
  \gdef\@pdfauthor{\infoauthors}
  \hypersetup{%
    pdfcreator={ceurart.cls},
    linkcolor={hscolor},
    urlcolor={hscolor},
    citecolor={hscolor},
    filecolor={hscolor},
    menucolor={hscolor},
  }
}
\makeatother

\begin{document}

\copyrightyear{2022}
\copyrightclause{Copyright for this paper by its authors.
   Use permitted under Creative Commons License Attribution 4.0
   International (CC BY 4.0).}
\conference{PAAR 2022: 8th Workshop on Practical Aspects of Automated
  Reasoning, August 11-12, 2022, Haifa, Israel}

\title{Generating Compressed Combinatory Proof Structures -- An Approach
  to Automated First-Order Theorem Proving}

\author{Christoph Wernhard}[%
  email=info@christophwernhard.com]
\address{University of Potsdam, Germany}

\begin{abstract}
Representing a proof tree by a combinator term that reduces to the tree lets
subtle forms of duplication within the tree materialize as duplicated subterms
of the combinator term. In a DAG representation of the combinator term these
straightforwardly factor into shared subgraphs.
To search for proofs, combinator terms can be enumerated, like clausal
tableaux, interwoven with unification of formulas that are associated with
nodes of the enumerated structures.
To restrict the search space, the enumeration can be based on proof schemas
defined as parameterized combinator terms.
We introduce here this ``combinator term as proof structure'' approach to
automated first-order proving, present an implementation and first
experimental results.
The approach builds on a term view of proof structures rooted in condensed
detachment and the connection method. It realizes features known from the
connection structure calculus, which has not been implemented so far.
\end{abstract}


\maketitle

\section{Introduction}
\label{sec-intro}

Goal-driven first-order provers such as \leanCoP \cite{leancop}, \name{PTTP}
\cite{pttp}, \SETHEO \cite{setheo:92} and \CMProver \cite{cw:pie:2016}, which
can be described as based on clausal tableaux \cite{letz:habil}, the
connection method \cite{bibel:atp:1982,bibel:otten:2020} or model elimination
\cite{loveland:1978}, in essence enumerate tree-shaped proof structures,
interwoven with unification of formulas that are associated with nodes of the
structures. While such provers do not compete with state-of-the art systems in
the range of solvable problems, they have merits that are relevant in certain
contexts: Proofs are typically emitted as data structures of simple and
detailed forms, making them suitable as inputs for further processing. The
provers facilitate comparing alternate proofs of a problem or influencing the
shape of proofs. Interpolation is an example where such features are quite
useful \cite{toman:2015:tableaux,cw:interpolation:2021}. The goal-driven mode
of operation and constraining the proof shape can be stretched to using such a
prover as processor of a full-fledged programming language (Prolog). In
general, through iterative deepening the emitted proofs tend to be short,
which again is useful for further processing, including integration with other
systems and presentation for humans. Implementations following the approach
are typically manageable and small (with \leanCoP outbidding all others
\cite{otten:2003:leancopinabstract}), making them attractive for adaptation to
specific logics \cite{otten:2014:mleancop,otten:2016:intui,otten:2021:nanocop}
and novel combinations with other techniques
\cite{kaliszyk:2015:femalecop,zombori:2020:prolog,zombori:2021:entropy,lazycop:2021,satcop:2021,faerber:2021:mlct}.
Another aspect of the approach with potential long-term relevance is its role
as a foundation for systematic investigations of first-order ATP, as for
example in
\cite{bibel:atp:1982,eder:relative:1992,letz:habil,cwwb:lukas:2021}.

Here we aim to preserve the merits of the clausal tableau or conventional
connection approach, with respect to theory as well as practice, while moving
on to stronger proving capabilities. In particular, we address a crucial
limitation of the conventional techniques: the restriction to tree structures
in contrast to DAG structures, where nodes with multiple incoming edges
correspond to multiply used lemmas. The issue has three aspects: First, DAGs
permit to share duplicated subtrees where the conventional handling of
variables as rigid (their scope is the whole structure) is not sufficient,
because, in general, each use of a lemma takes it in a fresh copy. Second, for
DAG enumeration a naturally adequate depth measure for iterative deepening is
the DAG size, i.e., the number of inner nodes of the DAG, in contrast to
conventionally used measures such as the number of tree nodes or tree height.

The third aspect is to extend the forms of duplication in the proof structure
that materialize as duplicated subtrees and are thus shareable in the DAG
representation. As an example of this aspect, consider the proof tree
\begin{equation}
  \label{eq-intro-tree}
  2 (2 (2 (2 (2 (2 (2 (2 1))))))).
\end{equation}
If leaf label $1$ represents the axiom $\fp(\fa)$, leaf label $2$ the axiom
$\fp(x) \imp \fp(\ff(x))$ and, given a proof $d_1$ of $P \imp Q$ and a proof
$d_2$ of $R$, the juxtaposition $d_1 d_2$ represents an inference step by
modus ponens with unification, i.e., $d_1 d_2$ is a proof of $Q\sigma$, where
$\sigma$ is the most general unifier of~$P$ and~$R$, then
(\ref{eq-intro-tree}) is a proof of
$\fp(\ff(\ff(\ff(\ff(\ff(\ff(\ff(\ff(\fa)))))))))$. Although in
(\ref{eq-intro-tree}) intuitively the repeated use of axiom~$2$ constitutes
some kind of duplication, there are no duplicated subtrees
in~(\ref{eq-intro-tree}), except of leaves. The minimal DAG of
(\ref{eq-intro-tree}) is identical to the tree. The intuitively observed form
of duplication is not captured by duplicated subtrees that could be factored
in the DAG representation.

The idea is now to use combinator terms as representations of proof
structures, which enrich the possibilities of sharing substructures. A
combinator is technically a $\lambda$-term without free variables. Combinators
form the basis of combinatory logic, introduced in the 1920s by Moses
Schönfinkel \cite{schoenfinkel:24:bausteine}, developed further by Haskell B.
Curry \cite{curry:1958}, and later used for functional programming languages
\cite{peytonjones:87} and for an empirical computation theory
\cite{wolfram:2021} (see \cite{wolfram:biblio:2021} for a comprehensive
bibliography). As common in the literature, we notate function application,
which now generalizes the modus ponens inference steps considered before, by
left associative juxtaposition. The $\c{B}$ combinator, for example, which can
be characterized by the equivalence
\begin{equation}
  \label{eq-protodef-b}
  \c{B} x y z \equiv x (y z).
\end{equation}
allows to represent~(\ref{eq-intro-tree}) by the combinator term
\begin{equation}
  \label{eq-intro-dag}
  \c{B} (\c{B} 2 2) (\c{B} 2 2) (\c{B} (\c{B} 2 2) (\c{B} 2 2) 1).
\end{equation}
Eliminating in (\ref{eq-intro-dag}) the occurrences of $\c{B}$ by rewriting
with (\ref{eq-protodef-b}) results in~(\ref{eq-intro-tree}). In
(\ref{eq-intro-dag}) the subtrees $\c{B} 2 2$ and $\c{B} (\c{B} 2 2) (\c{B} 2
2)$ occur multiply such that each of them has two incoming edges in the
minimal DAG of (\ref{eq-intro-dag}) (Fig.~\ref{fig-f8-b-dag-intro}). The
number of inner nodes of that DAG is only~6, compared to~8
for~(\ref{eq-intro-tree}).

\begin{wrapfigure}[18]{r}{4.0cm}
  \sffamily\small
  \centering
  \vspace{-12pt}  
  \caption{Term (\ref{eq-intro-dag}) as DAG.}
  \vspace{3pt}
  \label{fig-f8-b-dag-intro}
  {\normalfont \scalebox{0.4}{\begindg
  \pgfsetlinewidth{1bp}
\pgfsetcolor{black}
  \draw [->] (64.0bp,143.83bp) .. controls (64.0bp,136.13bp) and (64.0bp,126.97bp)  .. (64.0bp,108.41bp);
  \draw [->] (69.568bp,432.57bp) .. controls (72.223bp,424.26bp) and (75.46bp,414.12bp)  .. (81.484bp,395.27bp);
  \draw [->] (61.134bp,432.05bp) .. controls (57.208bp,407.48bp) and (50.12bp,363.1bp)  .. (43.903bp,324.17bp);
  \draw [->] (77.245bp,362.73bp) .. controls (71.146bp,353.18bp) and (63.162bp,340.69bp)  .. (50.741bp,321.25bp);
  \draw [->] (43.866bp,288.05bp) .. controls (47.792bp,263.48bp) and (54.88bp,219.1bp)  .. (61.097bp,180.17bp);
  \draw [->] (19.587bp,215.96bp) .. controls (22.898bp,178.31bp) and (30.562bp,91.187bp)  .. (35.404bp,36.145bp);
  \draw [->] (73.624bp,146.55bp) .. controls (79.665bp,136.08bp) and (87.038bp,121.69bp)  .. (91.0bp,108.0bp) .. controls (96.81bp,87.919bp) and (99.012bp,64.385bp)  .. (100.1bp,36.295bp);
  \draw [->] (27.755bp,218.73bp) .. controls (33.854bp,209.18bp) and (41.838bp,196.69bp)  .. (54.259bp,177.25bp);
  \draw [->] (89.019bp,359.83bp) .. controls (89.874bp,352.13bp) and (90.892bp,342.97bp)  .. (92.954bp,324.41bp);
  \draw [->] (57.601bp,72.937bp) .. controls (54.555bp,64.814bp) and (50.835bp,54.892bp)  .. (43.825bp,36.199bp);
  \draw [->] (35.432bp,288.57bp) .. controls (32.777bp,280.26bp) and (29.54bp,270.12bp)  .. (23.516bp,251.27bp);
  \draw [->] (72.169bp,73.662bp) .. controls (76.37bp,65.26bp) and (81.601bp,54.797bp)  .. (90.979bp,36.042bp);
\begin{scope}
  \definecolor{strokecol}{rgb}{0.0,0.0,0.0}
  \pgfsetstrokecolor{strokecol}
  \draw (18.0bp,234.0bp) ellipse (18.0bp and 18.0bp);
  \draw (18.0bp,234.0bp) node[font=\LARGE] {};
\end{scope}
\begin{scope}
  \definecolor{strokecol}{rgb}{0.0,0.0,0.0}
  \pgfsetstrokecolor{strokecol}
  \draw (64.0bp,90.0bp) ellipse (18.0bp and 18.0bp);
  \draw (64.0bp,90.0bp) node[font=\LARGE] {};
\end{scope}
\begin{scope}
  \definecolor{strokecol}{rgb}{0.0,0.0,0.0}
  \pgfsetstrokecolor{strokecol}
  \draw (87.0bp,378.0bp) ellipse (18.0bp and 18.0bp);
  \draw (87.0bp,378.0bp) node[font=\LARGE] {};
\end{scope}
\begin{scope}
  \definecolor{strokecol}{rgb}{0.0,0.0,0.0}
  \pgfsetstrokecolor{strokecol}
  \draw (41.0bp,306.0bp) ellipse (18.0bp and 18.0bp);
  \draw (41.0bp,306.0bp) node[font=\LARGE] {};
\end{scope}
\begin{scope}
  \definecolor{strokecol}{rgb}{0.0,0.0,0.0}
  \pgfsetstrokecolor{strokecol}
  \draw (64.0bp,162.0bp) ellipse (18.0bp and 18.0bp);
  \draw (64.0bp,162.0bp) node[font=\LARGE] {};
\end{scope}
\begin{scope}
  \definecolor{strokecol}{rgb}{0.0,0.0,0.0}
  \pgfsetstrokecolor{strokecol}
  \draw (64.0bp,450.0bp) ellipse (18.0bp and 18.0bp);
  \draw (64.0bp,450.0bp) node[font=\LARGE] {};
\end{scope}
\begin{scope}
  \definecolor{strokecol}{rgb}{0.0,0.0,0.0}
  \pgfsetstrokecolor{strokecol}
  \draw (113.0bp,324.0bp) -- (77.0bp,324.0bp) -- (77.0bp,288.0bp) -- (113.0bp,288.0bp) -- cycle;
  \draw (95.0bp,306.0bp) node[font=\LARGE] {1};
\end{scope}
\begin{scope}
  \definecolor{strokecol}{rgb}{0.0,0.0,0.0}
  \pgfsetstrokecolor{strokecol}
  \draw (55.0bp,36.0bp) -- (19.0bp,36.0bp) -- (19.0bp,0.0bp) -- (55.0bp,0.0bp) -- cycle;
  \draw (37.0bp,18.0bp) node[font=\LARGE] {$\c{B}$};
\end{scope}
\begin{scope}
  \definecolor{strokecol}{rgb}{0.0,0.0,0.0}
  \pgfsetstrokecolor{strokecol}
  \draw (118.0bp,36.0bp) -- (82.0bp,36.0bp) -- (82.0bp,0.0bp) -- (118.0bp,0.0bp) -- cycle;
  \draw (100.0bp,18.0bp) node[font=\LARGE] {2};
\end{scope}
\end{tikzpicture}}}
\end{wrapfigure}

For \emph{proof search}, terms such as (\ref{eq-intro-dag}), that is, proof
structures with combinators and identifiers of proper axioms\footnote{A
  \defname{proper axiom} is an axiom from the application problem, in contrast
  to the logical machinery.} as constants, can be \emph{enumerated}, just like
clausal tableaux. The interwoven formula unification is constrained from the
root by the proof goal, from the leaves by copies of axioms and from within
the structure by constraints induced by the modus ponens inference rule. A
combinator is handled just as an axiom with an implicational formula that is
associated in a specific way with its definition, the \name{principal type}
\cite{hindley:book:1997} of the combinator.

The proof search is then actually upon compressed structures. On the one hand,
these may be shorter than expanded structures but on the other hand, the
search space may also be increased because a single expanded structure can be
represented by many different compressed structures. For example,
(\ref{eq-intro-tree}), an expanded structure because it does not involve a
combinator, can be represented in compressed form with the $\c{B}$ combinator
not just by (\ref{eq-intro-dag}), but also by five further structures with the
same size of the minimal DAG, including, e.g., $\c{B} 2 2 (\c{B} 2 2 (\c{B} 2
2 (\c{B} 2 2 1)))$.

The balancing between short proofs and the vast number of compression
possibilities in proof search with compressed structures is addressed here
with a specific form of \name{proof schemas}, patterns, consisting of a
function symbol with arguments and an associated defining combinator term. If
search is performed with a specified set such proof schemas, the proof
structures are constructed only from constants representing proper axioms and
the function patterns. In proofs, instances of the function patterns can be
rewritten into their defining combinator terms with at most a linear increase
of the DAG size, making the proofs independent from possibly ad-hoc schemas
used for search.

In brief, we introduce an approach to first-order theorem proving that has its
background in goal-driven clausal tableaux or conventional realizations of the
connection method, but differently from these, is characterized by search over
combinatory compressed proof structures, with DAGs instead of trees as basic
structure for re-usable lemmas and DAG size as basic measure for iterative
deepening. A proof schema mechanism based on combinator terms is provided to
control search in presence of compression.

In Sect.~\ref{sec-outline} we flesh out the approach and sketch features of
\CCS, an implemented experimental prover. We then present in
Sect.~\ref{sec-experiments} empirical results, so far for the problems in the
\TPTP \cite{tptp} that are directly expressed using modus ponens and
unification, i.e., that are condensed detachment \cite{ulrich:legacy:2001}
problems. In Sect.~\ref{sec-conclusion} we conclude with pointers to related
work , next steps, open issues and perspectives. The implemented prover and
auxiliary tools are components of the latest version of the \CDTOOLS system
\cite{cw:cdtools}, available as free software from\par
     {\centering \vspace{4pt}
       \url{http://cs.christophwernhard.com/cdtools/}.\par \vspace{4pt}}
\noindent
That website also provides supplementary result tables, including graphical
proof visualizations, as well as detailed outputs from the experiments
described in the paper.

\section{Outline of the Approach}
\label{sec-outline}

\subsection{Tree/Term/DAG Representation of Condensed Detachment Proofs}
\label{sec-intro-mgt}

As basic inference mechanism we consider condensed detachment
\cite{prior:logicians:1956,kalman:cd:1983,hindley:meredith:cd:1990,ulrich:legacy:2001,cwwb:lukas:2021},
in other words modus ponens with unification, originally due to Carew A.
Meredith. An inference step by condensed detachment can be informally
described as deriving from major premise $P \imp Q$ and minor premise $R$ the
conclusion $Q$ under the most general unifier of $P$ and $R$.\footnote{For a
  recent comprehensive account of condensed detachment that relates to the
  connection method see \cite{cwwb:lukas:2021}.} The structural
component of a condensed detachment proof is a full binary tree where the left
child of an inner node represents the subproof of the major premise and the
right child that of the minor premise. Leaves represent axioms. Such a tree
can be conveniently written as a so called D-term, a term with the binary
function symbol $\D$ for inner nodes and constants identifying axioms as
leaves. The following term provides an example. It tree visualization is shown
in Fig.~\ref{fig-dterm-dag}(a).
\begin{equation}
  \label{eq-dterm}
   \D(\D(1, 1), \D(\D(1, \D(1, 1)), \D(1, \D(1, 1)))).
\end{equation}
For given axioms associated with the constants, the ``most general'' formula
proven by a D-term, called its \name{most general theorem (MGT)}
\cite{cwwb:lukas:2021} (or \name{principal type}
\cite{hindley:meredith:cd:1990,hindley:book:1997}), can be determined through
unification as follows: The MGT of leaf, a constant D-term, is a fresh
instance of the axiom labeled by the constant. If $P \imp Q$ and $R$ are the
MGTs of D-terms~$d_1$ and~$d_2$, respectively, then the MGT of $\D(d_1,d_1)$
is $Q\mgu(\{P,R\})$.\footnote{We use the common postfix notation for
  application of a substitution. If $M$ is a set of pairs of terms that has a
  unifier, then $\mgu(M)$ denotes the most general unifier of $M$. If $M$
  contains just a single pair $\{t,u\}$, we use $\mgu(\{t,u\})$ as shorthand
  for $\mgu(\{\{t,u\}\})$. For a precise account of condensed detachment, we
  assume w.l.o.g. that $\mgu(M)$ has the \name{clean} property
  \cite{cwwb:lukas:2021}.} If there is no substitution that satisfies these
unification constraints for all nodes of a D-term, its MGT is said to be
\emph{not defined}.

Notice that each instance of an axiom, associated with a leaf, is before
consideration of the unification constraints a fresh copy of the axiom. By the
unification constraints, variables are propagated throughout the tree. Hence
the variables are rigid, that is, their scope comprises all formulas
associated with nodes of the whole structure. For the example axioms in
Table~\ref{tab-fn-axioms} it is easy to see that the MGT of the D-term~$1$ is
$\P(\fa)$, the MGT of $2$ is $\P(x) \imp \P(\ff(x))$ (modulo renaming $x$ to a
fresh variable), the MGT of $\D(2,1)$ is $\P(\ff(\fa))$, and the MGT of
$\D(2,\D(2,1))$ is $\P(\ff(\ff(\fa)))$.

\begin{table}[h]
  \centering
  \caption{Examples of proper axioms with numbers as identifying constants.}
  \label{tab-fn-axioms}
$\begin{array}{cl}
\text{Axiom ID} & \text{Axiom formula}\\\midrule
1 & \P(\fa)\\
2 & \P(x) \imp \P(\ff(x))\\
\end{array}$
\end{table}

$\D$ is the only non-constant function symbol used in D-terms. Moreover, a
condensed detachment step with premises $P \imp Q$ and $R$ may be viewed as
\emph{application} of the function $\lambda x. Q\mgu(\{P,x\})$ to $R$. This
suggests to present complex D-terms in the common notation for application in
$\lambda$-terms and terms of combinatory logic. For example,
$\D(\D(1,2),\D(3,4))$ is then written as $1 2 (3 4)$, and~(\ref{eq-dterm}) as
\begin{equation}
  \label{eq-dterm-sk}
  1 1 (1 (1 1) (1 (1 1))).
\end{equation}

\begin{figure}
  \centering
  \caption{D-term (\ref{eq-dterm}), shown in applicative notation in
    (\ref{eq-dterm-sk}), as tree~(a) and as DAG~(b). In~(b) the labels of
    inner nodes in~(b) show the corresponding factor labels
    from~(\ref{eq-dterm-dag}).}
  \label{fig-dterm-dag}
  \hspace*{\fill}
  \raisebox{4.3cm}{(a)}
  {\normalfont \scalebox{0.4}{\begindg
  \pgfsetlinewidth{1bp}
\pgfsetcolor{black}
  \draw [->] (126.0bp,215.83bp) .. controls (126.0bp,208.13bp) and (126.0bp,198.97bp)  .. (126.0bp,180.41bp);
  \draw [->] (105.4bp,288.94bp) .. controls (108.59bp,280.42bp) and (112.53bp,269.92bp)  .. (119.65bp,250.94bp);
  \draw [->] (132.4bp,72.937bp) .. controls (135.44bp,64.814bp) and (139.17bp,54.892bp)  .. (146.18bp,36.199bp);
  \draw [->] (61.072bp,219.43bp) .. controls (54.361bp,210.48bp) and (45.566bp,198.75bp)  .. (31.596bp,180.13bp);
  \draw [->] (180.0bp,143.83bp) .. controls (180.0bp,136.13bp) and (180.0bp,126.97bp)  .. (180.0bp,108.41bp);
  \draw [->] (119.6bp,72.937bp) .. controls (116.56bp,64.814bp) and (112.83bp,54.892bp)  .. (105.82bp,36.199bp);
  \draw [->] (115.07bp,147.43bp) .. controls (108.36bp,138.48bp) and (99.566bp,126.75bp)  .. (85.596bp,108.13bp);
  \draw [->] (230.54bp,72.202bp) .. controls (229.0bp,64.3bp) and (227.16bp,54.811bp)  .. (223.52bp,36.093bp);
  \draw [->] (242.88bp,74.021bp) .. controls (247.59bp,65.544bp) and (253.5bp,54.896bp)  .. (263.88bp,36.218bp);
  \draw [->] (126.0bp,143.83bp) .. controls (126.0bp,136.13bp) and (126.0bp,126.97bp)  .. (126.0bp,108.41bp);
  \draw [->] (92.601bp,288.94bp) .. controls (89.407bp,280.42bp) and (85.471bp,269.92bp)  .. (78.353bp,250.94bp);
  \draw [->] (136.93bp,219.43bp) .. controls (144.39bp,209.48bp) and (154.44bp,196.08bp)  .. (169.09bp,176.55bp);
  \draw [->] (190.93bp,147.43bp) .. controls (198.39bp,137.48bp) and (208.44bp,124.08bp)  .. (223.09bp,104.55bp);
  \draw [->] (72.0bp,215.83bp) .. controls (72.0bp,208.13bp) and (72.0bp,198.97bp)  .. (72.0bp,180.41bp);
\begin{scope}
  \definecolor{strokecol}{rgb}{0.0,0.0,0.0}
  \pgfsetstrokecolor{strokecol}
  \draw (117.0bp,36.0bp) -- (81.0bp,36.0bp) -- (81.0bp,0.0bp) -- (117.0bp,0.0bp) -- cycle;
  \draw (99.0bp,18.0bp) node[font=\LARGE] {1};
\end{scope}
\begin{scope}
  \definecolor{strokecol}{rgb}{0.0,0.0,0.0}
  \pgfsetstrokecolor{strokecol}
  \draw (238.0bp,36.0bp) -- (202.0bp,36.0bp) -- (202.0bp,0.0bp) -- (238.0bp,0.0bp) -- cycle;
  \draw (220.0bp,18.0bp) node[font=\LARGE] {1};
\end{scope}
\begin{scope}
  \definecolor{strokecol}{rgb}{0.0,0.0,0.0}
  \pgfsetstrokecolor{strokecol}
  \draw (234.0bp,90.0bp) ellipse (18.0bp and 18.0bp);
\end{scope}
\begin{scope}
  \definecolor{strokecol}{rgb}{0.0,0.0,0.0}
  \pgfsetstrokecolor{strokecol}
  \draw (180.0bp,162.0bp) ellipse (18.0bp and 18.0bp);
\end{scope}
\begin{scope}
  \definecolor{strokecol}{rgb}{0.0,0.0,0.0}
  \pgfsetstrokecolor{strokecol}
  \draw (126.0bp,90.0bp) ellipse (18.0bp and 18.0bp);
\end{scope}
\begin{scope}
  \definecolor{strokecol}{rgb}{0.0,0.0,0.0}
  \pgfsetstrokecolor{strokecol}
  \draw (126.0bp,162.0bp) ellipse (18.0bp and 18.0bp);
\end{scope}
\begin{scope}
  \definecolor{strokecol}{rgb}{0.0,0.0,0.0}
  \pgfsetstrokecolor{strokecol}
  \draw (36.0bp,180.0bp) -- (0.0bp,180.0bp) -- (0.0bp,144.0bp) -- (36.0bp,144.0bp) -- cycle;
  \draw (18.0bp,162.0bp) node[font=\LARGE] {1};
\end{scope}
\begin{scope}
  \definecolor{strokecol}{rgb}{0.0,0.0,0.0}
  \pgfsetstrokecolor{strokecol}
  \draw (72.0bp,234.0bp) ellipse (18.0bp and 18.0bp);
\end{scope}
\begin{scope}
  \definecolor{strokecol}{rgb}{0.0,0.0,0.0}
  \pgfsetstrokecolor{strokecol}
  \draw (126.0bp,234.0bp) ellipse (18.0bp and 18.0bp);
\end{scope}
\begin{scope}
  \definecolor{strokecol}{rgb}{0.0,0.0,0.0}
  \pgfsetstrokecolor{strokecol}
  \draw (292.0bp,36.0bp) -- (256.0bp,36.0bp) -- (256.0bp,0.0bp) -- (292.0bp,0.0bp) -- cycle;
  \draw (274.0bp,18.0bp) node[font=\LARGE] {1};
\end{scope}
\begin{scope}
  \definecolor{strokecol}{rgb}{0.0,0.0,0.0}
  \pgfsetstrokecolor{strokecol}
  \draw (198.0bp,108.0bp) -- (162.0bp,108.0bp) -- (162.0bp,72.0bp) -- (198.0bp,72.0bp) -- cycle;
  \draw (180.0bp,90.0bp) node[font=\LARGE] {1};
\end{scope}
\begin{scope}
  \definecolor{strokecol}{rgb}{0.0,0.0,0.0}
  \pgfsetstrokecolor{strokecol}
  \draw (90.0bp,180.0bp) -- (54.0bp,180.0bp) -- (54.0bp,144.0bp) -- (90.0bp,144.0bp) -- cycle;
  \draw (72.0bp,162.0bp) node[font=\LARGE] {1};
\end{scope}
\begin{scope}
  \definecolor{strokecol}{rgb}{0.0,0.0,0.0}
  \pgfsetstrokecolor{strokecol}
  \draw (171.0bp,36.0bp) -- (135.0bp,36.0bp) -- (135.0bp,0.0bp) -- (171.0bp,0.0bp) -- cycle;
  \draw (153.0bp,18.0bp) node[font=\LARGE] {1};
\end{scope}
\begin{scope}
  \definecolor{strokecol}{rgb}{0.0,0.0,0.0}
  \pgfsetstrokecolor{strokecol}
  \draw (99.0bp,306.0bp) ellipse (18.0bp and 18.0bp);
\end{scope}
\begin{scope}
  \definecolor{strokecol}{rgb}{0.0,0.0,0.0}
  \pgfsetstrokecolor{strokecol}
  \draw (90.0bp,108.0bp) -- (54.0bp,108.0bp) -- (54.0bp,72.0bp) -- (90.0bp,72.0bp) -- cycle;
  \draw (72.0bp,90.0bp) node[font=\LARGE] {1};
\end{scope}
\end{tikzpicture}}}
  \hspace*{\fill}
  \raisebox{4.3cm}{(b)\hspace*{1em}}
  {\normalfont \scalebox{0.4}{\begindg
  \pgfsetlinewidth{1bp}
\pgfsetcolor{black}
  \draw [->] (23.568bp,288.57bp) .. controls (26.223bp,280.26bp) and (29.46bp,270.12bp)  .. (35.484bp,251.27bp);
  \draw [->] (35.16bp,72.937bp) .. controls (34.317bp,64.987bp) and (34.059bp,55.313bp)  .. (35.045bp,36.199bp);
  \draw [->] (46.84bp,72.937bp) .. controls (47.683bp,64.987bp) and (47.941bp,55.313bp)  .. (46.955bp,36.199bp);
  \draw [->] (41.0bp,143.83bp) .. controls (41.0bp,136.13bp) and (41.0bp,126.97bp)  .. (41.0bp,108.41bp);
  \draw [->] (30.923bp,146.67bp) .. controls (24.691bp,136.25bp) and (17.275bp,121.87bp)  .. (14.0bp,108.0bp) .. controls (10.324bp,92.428bp) and (10.324bp,87.572bp)  .. (14.0bp,72.0bp) .. controls (16.17bp,62.809bp) and (20.156bp,53.4bp)  .. (29.249bp,36.176bp);
  \draw [->] (35.16bp,216.94bp) .. controls (34.297bp,208.81bp) and (34.048bp,198.88bp)  .. (35.121bp,179.44bp);
  \draw [->] (46.84bp,216.94bp) .. controls (47.703bp,208.81bp) and (47.952bp,198.88bp)  .. (46.879bp,179.44bp);
  \draw [->] (14.651bp,288.2bp) .. controls (9.6186bp,258.04bp) and (1.9127bp,195.2bp)  .. (14.0bp,144.0bp) .. controls (16.431bp,133.7bp) and (21.142bp,123.13bp)  .. (30.923bp,105.33bp);
\begin{scope}
  \definecolor{strokecol}{rgb}{0.0,0.0,0.0}
  \pgfsetstrokecolor{strokecol}
  \draw (59.0bp,36.0bp) -- (23.0bp,36.0bp) -- (23.0bp,0.0bp) -- (59.0bp,0.0bp) -- cycle;
  \draw (41.0bp,18.0bp) node[font=\LARGE] {1};
\end{scope}
\begin{scope}
  \definecolor{strokecol}{rgb}{0.0,0.0,0.0}
  \pgfsetstrokecolor{strokecol}
  \draw (41.0bp,234.0bp) ellipse (18.0bp and 18.0bp);
  \draw (41.0bp,234.0bp) node[font=\LARGE] {};
\end{scope}
\begin{scope}
  \definecolor{strokecol}{rgb}{0.0,0.0,0.0}
  \pgfsetstrokecolor{strokecol}
  \draw (41.0bp,162.0bp) ellipse (18.0bp and 18.0bp);
  \draw (41.0bp,162.0bp) node[font=\LARGE] {3};
\end{scope}
\begin{scope}
  \definecolor{strokecol}{rgb}{0.0,0.0,0.0}
  \pgfsetstrokecolor{strokecol}
  \draw (41.0bp,90.0bp) ellipse (18.0bp and 18.0bp);
  \draw (41.0bp,90.0bp) node[font=\LARGE] {2};
\end{scope}
\begin{scope}
  \definecolor{strokecol}{rgb}{0.0,0.0,0.0}
  \pgfsetstrokecolor{strokecol}
  \draw (18.0bp,306.0bp) ellipse (18.0bp and 18.0bp);
  \draw (18.0bp,306.0bp) node[font=\LARGE] {4};
\end{scope}
\end{tikzpicture}}}
  \hspace*{\fill}          
\end{figure}

Useful measures for the size of a D-term are \name{tree size}, i.e., the
number of inner nodes, \name{height}, i.e., the height of the tree, and
\name{compacted size}, i.e., the number of inner nodes of the minimal DAG
representing the tree, or, equivalently, the number of distinct compound
subterms. For example, the D-term shown in (\ref{eq-dterm}),
(\ref{eq-dterm-sk}) and Fig.~\ref{fig-dterm-dag} has tree size~7, height~4 and
compacted size~4. A DAG can be represented textually in efficient form by a
list of factors, paired with labels to express references. The following list
of factors gives an example for the minimal DAG of (\ref{eq-dterm-sk}) as
shown in Fig.~\ref{fig-dterm-dag}.(b).
\begin{equation}
  \label{eq-dterm-dag}
  [2 = 1 1,\; 3 = 1 2,\; 4 = 2 (3 3)].
\end{equation}
In this example numbers are used as factor labels, starting from $2$, since
$1$ was the largest number used as axiom identifier. The presentation order
mimics building up the proof by applying condensed detachment to previously
obtained lemmas, with the axiom identifiers, just $1$ in the example,
presupposed. The largest number, $4$ in the example, corresponds to the root
of the term. The original expanded D-term, (\ref{eq-dterm-sk}) in the example,
can be obtained from the root factor, $2(33)$ in the example, by exhaustively
replacing labels with their designated factor. If the factor list represents
the \emph{minimal} DAG, the compacted size of the represented D-term can be
read off by counting the occurrences of application on the right hand sides.

\subsection{Combinatory Compression}

As running example, we consider proving $\P(\ff^{8}(\fa))$, which stands for
$\P(\ff(\ff(\ff(\ff(\ff(\ff(\ff(\ff(\fa)))))))))$, from the axioms in
Table~\ref{tab-fn-axioms}.\footnote{This example, already sketched in the
  introduction, is from \cite[Sects.~2.10 and~3.7]{bibel:eder:1993}, where it
  is used to explicate the connection structure calculus. It was chosen here
  to indicate parallels to the connection structure calculus.}
There is a single D-term which proves this, namely
\begin{equation}
  \label{eq-f8-dterm}
  2 (2 (2 (2 (2 (2 (2 (2 1))))))).
\end{equation}
Its tree size as well as its compacted size are both~$8$. As indicated by the
compacted size, it has $8$ distinct compound subterms: $2 1$, $2 (2 1)$, $2 (2
(2 1))$, $\ldots$, $2 (2 (2 (2 (2 (2 (2 (2 1)))))))$, each of them with just a
single occurrence. Hence its minimal DAG is identical to the tree. That
axiom~$2$ is applied eight times is not reflected in any multiply occurring
compound subterm. This can be remedied with the $\c{B}$ combinator, defined as
\begin{equation}
  \label{eq-b-def-lambda}
  \c{B}\; \eqdef\; \lambda xyz \ldot x (y z).
\end{equation}
Occurrences of $\c{B}$ where it is applied to three arguments, matching the
number of arguments of its defining $\lambda$-term, can be eliminated by
rewriting to the body of that $\lambda$-term, i.e., with the
equivalence-preserving rewrite rule
\begin{equation}
  \label{eq-rule-b}
  \c{B} x y z \rimp x (y z).
\end{equation}

We now generalize D-terms to permit as constants, in addition to the
identifiers of proper axioms, also combinators. To emphasize that no
combinators occur in a D-term, we call it a \defname{pure D-term}. To
emphasize that combinators are permitted, we call it \defname{CL-term}, with
CL suggesting \name{combinatory logic}. If only combinators, in contrast to
identifiers of proper axioms, occur as constants, we speak of a \defname{pure
  CL-term}. For determining the MGT, a combinator is just considered as
identifier of an associated axiom, determined as the principal type of the
combinator.\footnote{For examples, see Table~\ref{tab-combinators} below.
  In \label{foot-prinicipal-type} the literature there are different
  approaches to formally deal with the relationship of combinators and
  associated implicational formulas: In \cite[Table~3E2a]{hindley:book:1997}
  the formulas are the \name{principal types} of the combinators or equivalent
  $\lambda$-terms. In \cite[Sect.~9D]{curry:1958} they are called
  \name{functional characters of combinators}. In
  \cite[Chap.~6]{bimbo:combinatory:2012} they are used to specify operational
  models for an algebraic view on combinators. In \cite{rezus:1982} the
  formulas are taken as starting point.} We can now
express~(\ref{eq-f8-dterm}) as the CL-term
\begin{equation}
  \label{eq-f8-with-b}
\c{B} (\c{B} 2 2) (\c{B} 2 2) (\c{B} (\c{B} 2 2) (\c{B} 2 2) 1).
\end{equation}

\begin{wrapfigure}[18]{r}{4.3cm}
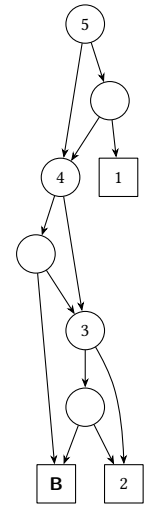

  \sffamily\small
  \centering
  \vspace{-12pt}  
  \caption{D-term (\ref{eq-f8-with-b}) as DAG.}
  \vspace{3pt}
  \label{fig-f8-b-dag}
  {\normalfont \scalebox{0.4}{\begindg
  \pgfsetlinewidth{1bp}
\pgfsetcolor{black}
  \draw [->] (64.0bp,143.83bp) .. controls (64.0bp,136.13bp) and (64.0bp,126.97bp)  .. (64.0bp,108.41bp);
  \draw [->] (69.568bp,432.57bp) .. controls (72.223bp,424.26bp) and (75.46bp,414.12bp)  .. (81.484bp,395.27bp);
  \draw [->] (61.134bp,432.05bp) .. controls (57.208bp,407.48bp) and (50.12bp,363.1bp)  .. (43.903bp,324.17bp);
  \draw [->] (77.245bp,362.73bp) .. controls (71.146bp,353.18bp) and (63.162bp,340.69bp)  .. (50.741bp,321.25bp);
  \draw [->] (43.866bp,288.05bp) .. controls (47.792bp,263.48bp) and (54.88bp,219.1bp)  .. (61.097bp,180.17bp);
  \draw [->] (19.587bp,215.96bp) .. controls (22.898bp,178.31bp) and (30.562bp,91.187bp)  .. (35.404bp,36.145bp);
  \draw [->] (73.624bp,146.55bp) .. controls (79.665bp,136.08bp) and (87.038bp,121.69bp)  .. (91.0bp,108.0bp) .. controls (96.81bp,87.919bp) and (99.012bp,64.385bp)  .. (100.1bp,36.295bp);
  \draw [->] (27.755bp,218.73bp) .. controls (33.854bp,209.18bp) and (41.838bp,196.69bp)  .. (54.259bp,177.25bp);
  \draw [->] (89.019bp,359.83bp) .. controls (89.874bp,352.13bp) and (90.892bp,342.97bp)  .. (92.954bp,324.41bp);
  \draw [->] (57.601bp,72.937bp) .. controls (54.555bp,64.814bp) and (50.835bp,54.892bp)  .. (43.825bp,36.199bp);
  \draw [->] (35.432bp,288.57bp) .. controls (32.777bp,280.26bp) and (29.54bp,270.12bp)  .. (23.516bp,251.27bp);
  \draw [->] (72.169bp,73.662bp) .. controls (76.37bp,65.26bp) and (81.601bp,54.797bp)  .. (90.979bp,36.042bp);
\begin{scope}
  \definecolor{strokecol}{rgb}{0.0,0.0,0.0}
  \pgfsetstrokecolor{strokecol}
  \draw (18.0bp,234.0bp) ellipse (18.0bp and 18.0bp);
  \draw (18.0bp,234.0bp) node[font=\LARGE] {};
\end{scope}
\begin{scope}
  \definecolor{strokecol}{rgb}{0.0,0.0,0.0}
  \pgfsetstrokecolor{strokecol}
  \draw (64.0bp,90.0bp) ellipse (18.0bp and 18.0bp);
  \draw (64.0bp,90.0bp) node[font=\LARGE] {};
\end{scope}
\begin{scope}
  \definecolor{strokecol}{rgb}{0.0,0.0,0.0}
  \pgfsetstrokecolor{strokecol}
  \draw (87.0bp,378.0bp) ellipse (18.0bp and 18.0bp);
  \draw (87.0bp,378.0bp) node[font=\LARGE] {};
\end{scope}
\begin{scope}
  \definecolor{strokecol}{rgb}{0.0,0.0,0.0}
  \pgfsetstrokecolor{strokecol}
  \draw (41.0bp,306.0bp) ellipse (18.0bp and 18.0bp);
  \draw (41.0bp,306.0bp) node[font=\LARGE] {4};
\end{scope}
\begin{scope}
  \definecolor{strokecol}{rgb}{0.0,0.0,0.0}
  \pgfsetstrokecolor{strokecol}
  \draw (64.0bp,162.0bp) ellipse (18.0bp and 18.0bp);
  \draw (64.0bp,162.0bp) node[font=\LARGE] {3};
\end{scope}
\begin{scope}
  \definecolor{strokecol}{rgb}{0.0,0.0,0.0}
  \pgfsetstrokecolor{strokecol}
  \draw (64.0bp,450.0bp) ellipse (18.0bp and 18.0bp);
  \draw (64.0bp,450.0bp) node[font=\LARGE] {5};
\end{scope}
\begin{scope}
  \definecolor{strokecol}{rgb}{0.0,0.0,0.0}
  \pgfsetstrokecolor{strokecol}
  \draw (113.0bp,324.0bp) -- (77.0bp,324.0bp) -- (77.0bp,288.0bp) -- (113.0bp,288.0bp) -- cycle;
  \draw (95.0bp,306.0bp) node[font=\LARGE] {1};
\end{scope}
\begin{scope}
  \definecolor{strokecol}{rgb}{0.0,0.0,0.0}
  \pgfsetstrokecolor{strokecol}
  \draw (55.0bp,36.0bp) -- (19.0bp,36.0bp) -- (19.0bp,0.0bp) -- (55.0bp,0.0bp) -- cycle;
  \draw (37.0bp,18.0bp) node[font=\LARGE] {$\c{B}$};
\end{scope}
\begin{scope}
  \definecolor{strokecol}{rgb}{0.0,0.0,0.0}
  \pgfsetstrokecolor{strokecol}
  \draw (118.0bp,36.0bp) -- (82.0bp,36.0bp) -- (82.0bp,0.0bp) -- (118.0bp,0.0bp) -- cycle;
  \draw (100.0bp,18.0bp) node[font=\LARGE] {2};
\end{scope}
\end{tikzpicture}}}
\end{wrapfigure}

\noindent
By exhaustively rewriting~(\ref{eq-f8-with-b}) with~(\ref{eq-rule-b}) we
obtain the pure D-term~(\ref{eq-f8-dterm}). We call the CL-term
(\ref{eq-f8-with-b}) a \defname{combinatory compression} of
(\ref{eq-f8-dterm}). Speaking of \name{compression} is justified, since the
(\ref{eq-f8-with-b}) has with~$6$ a smaller compacted size than
(\ref{eq-f8-dterm}), whose compacted size is~$8$. Figure~\ref{fig-f8-b-dag}
shows the minimal DAG of~(\ref{eq-f8-with-b}). As list of
factors it can be written as
\begin{equation}
  \label{eq-f8-b-dag}
  [3 = \c{B} 2 2,\;
   4 = \c{B} 3 3,\;
   5 = 4 (4 1)].
\end{equation}

While $\c{B}$ can be eliminated completely from the combinatory compressed
form~(\ref{eq-f8-with-b}) by rewriting with~(\ref{eq-rule-b}), the rewrite
rule can not be applied in any of the individual factors shown in
(\ref{eq-f8-b-dag}), because $\c{B}$ occurs in these only with two arguments.
The construction of~(\ref{eq-f8-with-b}) can be generalized to proofs of
$\P(\ff^{2^n}(\fa))$ for arbitrary $n \geq 1$. While the pure D-term then
always has compacted size~$2^n$, the compacted size of the CL-term with
$\c{B}$ is just~$2n$. For example, for $n=4$, to prove $\P(\ff^{16}(\fa))$,
the minimal DAG is $[3 = \c{B} 2 2,\; 4 = \c{B} 3 3,\; 5 = \c{B} 4 4,\; 6 = 5
  (5 1)]$.

\subsection{Proof Search with Compressed Combinatory Proof Structures}

Goal-driven clausal tableau provers enumerate proof structures, tableaux,
interwoven with unification of formulas associated with nodes of the
structures. These structures may be viewed as terms, pure D-terms in the
setting of condensed detachment. Now, like tableaux or pure D-terms, also
CL-terms can be enumerated as basis for first-order proof search. If we add
the combinator $\c{B}$ as axiom identifier that designates its principal type
(shown in Table~\ref{tab-combinators} below) to our proper axioms from
Table~\ref{tab-fn-axioms}, let $\P(\ff^8(\fa))$ be the goal, and enumerate
D-terms (or, more precisely, CL-terms, since now $\c{B}$ is permitted) by
iterative deepening upon compacted size, we find no proofs with compacted size
less than 6 and six proofs with compacted size~6, specifically:
\begin{equation}
  \label{eq-f8-b-proofs}
  \begin{array}{ll}
   \text{(a)} & [3 = \c{B} 2 2,\; 4 = 3 (3 (3 (3 1)))].\\
   \text{(b)} & [3 = \c{B} 2,\; 4 = 3 2,\; 5 = 3 4,\; 6 = 4 (5 (5 1))].\\
   \text{(c)} & [3 = \c{B} 2,\; 4 = 3 2,\; 5 = 3 4,\; 6 = 5 (5 (4 1))].\\
   \text{(d)} & [3 = \c{B} 2,\; 4 = 3 2,\; 5 = 3 4,\; 6 = 5 (4 (5 1))].\\
   \text{(e)} & [3 = \c{B} 2,\; 4 = 3 (3 (3 2)),\; 5 = 4 (4 1)].\\
   \text{(f)} & [3 = \c{B} 2 2,\; 4 = \c{B} 3 3,\; 5 = 4 (4 1)].\\
  \end{array}
\end{equation}
CL-term~(f) was shown above as~(\ref{eq-f8-with-b}). All six CL-terms reduce
with the rule~(\ref{eq-rule-b}) for $\c{B}$ to the same normal form,
(\ref{eq-f8-dterm}) and thus represent just different compressions of the same
``expanded'' proof, the pure D-term~(\ref{eq-f8-dterm}). For CL-terms~(a)
and~(f) the factors involve applications of $\c{B}$ to two arguments,
and~(b)--(e) to a single argument. Since the rewrite rule for $\c{B}$ is only
applicable in presence of three arguments, it fails to rewrite any individual
factor of these CL-terms.
Observing for the example that whether $\c{B}$ is applied in factors to one or
two arguments has no influence on the minimal proof size, we wish to restrict
the search space by constraining permitted occurrences of $\c{B}$ to the
two-argument cases. This can be achieved by using combinators not directly for
enumeration, but via parameterized \name{proof schemas}, whose semantics is
specified by a parameterized CL-term. Table~\ref{tab-f8-schemas} gives two
examples.
\begin{table}[h]
  \vspace{-5pt}
  \centering
  \caption{Proof schemas for use with the axioms from
    Table~\ref{tab-fn-axioms} and goals $\P(\ff^n(\fa))$.}
  \label{tab-f8-schemas}
$\begin{array}{llll}
  \text{\it Schema} & \text{\it CL-term} & \text{\it $\lambda$-term} &
  \text{\it Resolution-like view}\\\midrule
  \f{r}_0(p,q) & pq
  & pq
  & A \nimp B,\, D\; \derives\; B\mgu(\{A, D\}) \\
  \f{r}_1(p,q) & \c{B}pq
  & \lambda x. p(qx)
  & A \nimp B,\, C \nimp D\; \derives\; (C \nimp B)\mgu(\{A, D\}) \\
\end{array}$
\end{table}

A proof schema is determined by a \name{schema}, a pattern for use in the
construction of proof terms and a \name{defining CL-term} that specifies a
semantics. We call proof terms that are built-up from such schemas (and
possibly also application, combinators and identifiers of proper axioms)
\defname{PS-terms}. The $\lambda$-term in Table~\ref{tab-f8-schemas} is
equivalent to the defining CL-term and provides an alternate, sometimes more
intuitive, way to specify the semantics. The table also shows the effect of
the proof schema in terms of a binary resolution step between Horn clauses,
indicating a correspondence of proof schemas to restricted forms of lemma
computation by resolution.

\enlargethispage{15pt}

By enumerating PS-terms constructed from the schemas in
Table~\ref{tab-f8-schemas} and the identifiers~$1,2$ of proper axioms in
Table~\ref{tab-fn-axioms} with iterative deepening upon compacted size we find
the following two proofs with least compacted size, which is~4, shown also in
Fig.~\ref{fig-dterm-f8r}.
\begin{equation}
  \label{eq-f8r-ab}
  \begin{array}{ll}
  \text{(a)}   & [3 = \f{r}_1(2, 2),\; 4 = \f{r}_1(3, 3),\; 5 = \f{r}_0(4, \f{r}_0(4, 1))].\\
  \text{(b)} & [3 = \f{r}_1(2, 2),\; 4 = \f{r}_1(3, 3),\; 5 = \f{r}_0(\f{r}_1(4, 4), 1)].
  \end{array}
  \vspace{-1pt}
\end{equation}

\begin{wrapfigure}[12]{r}{5.4cm}
  \sffamily\small
  \centering
  \vspace{-12pt}
  \caption{Proofs (\ref{eq-f8r-ab}).(a) and (\ref{eq-f8r-ab}).(b).}
  \label{fig-dterm-f8r}
  \vspace{6pt}
  \raisebox{4.3cm}{(a)}
  {\normalfont \scalebox{0.4}{\begindg
  \pgfsetlinewidth{1bp}
\pgfsetcolor{black}
  \draw [->] (60.778bp,288.57bp) .. controls (63.987bp,280.32bp) and (67.894bp,270.27bp)  .. (75.187bp,251.52bp);
  \draw [->] (50.635bp,288.05bp) .. controls (46.027bp,263.48bp) and (37.706bp,219.1bp)  .. (30.407bp,180.17bp);
  \draw [->] (69.52bp,217.66bp) .. controls (62.488bp,208.46bp) and (53.565bp,196.78bp)  .. (39.431bp,178.27bp);
  \draw [->] (21.084bp,144.2bp) .. controls (20.28bp,136.18bp) and (20.057bp,126.52bp)  .. (21.105bp,107.59bp);
  \draw [->] (32.916bp,144.2bp) .. controls (33.72bp,136.18bp) and (33.943bp,126.52bp)  .. (32.895bp,107.59bp);
  \draw [->] (21.084bp,72.202bp) .. controls (20.3bp,64.386bp) and (20.069bp,55.017bp)  .. (21.055bp,36.093bp);
  \draw [->] (32.916bp,72.202bp) .. controls (33.7bp,64.386bp) and (33.931bp,55.017bp)  .. (32.945bp,36.093bp);
  \draw [->] (84.019bp,215.83bp) .. controls (84.874bp,208.13bp) and (85.892bp,198.97bp)  .. (87.954bp,180.41bp);
\begin{scope}
  \definecolor{strokecol}{rgb}{0.0,0.0,0.0}
  \pgfsetstrokecolor{strokecol}
  \draw (82.0bp,234.0bp) ellipse (27.0bp and 18.0bp);
  \draw (82.0bp,234.0bp) node[font=\LARGE] {\rnode{$\f{r}_0$}};
\end{scope}
\begin{scope}
  \definecolor{strokecol}{rgb}{0.0,0.0,0.0}
  \pgfsetstrokecolor{strokecol}
  \draw (27.0bp,162.0bp) ellipse (27.0bp and 18.0bp);
  \draw (27.0bp,162.0bp) node[font=\LARGE] {\rnode{$\f{r}_1$}};
  \draw (27.0bp-35bp,162.0bp+15bp) node[font=\Large] {$4$};   
\end{scope}
\begin{scope}
  \definecolor{strokecol}{rgb}{0.0,0.0,0.0}
  \pgfsetstrokecolor{strokecol}
  \draw (27.0bp,90.0bp) ellipse (27.0bp and 18.0bp);
  \draw (27.0bp,90.0bp) node[font=\LARGE] {\rnode{$\f{r}_1$}};
  \draw (27.0bp-35bp,90.0bp+15bp) node[font=\Large] {$3$};     
\end{scope}
\begin{scope}
  \definecolor{strokecol}{rgb}{0.0,0.0,0.0}
  \pgfsetstrokecolor{strokecol}
  \draw (54.0bp,306.0bp) ellipse (27.0bp and 18.0bp);
  \draw (54.0bp,306.0bp) node[font=\LARGE] {\rnode{$\f{r}_0$}};
  \draw (54.0bp-35bp,306.0bp+15bp) node[font=\Large] {$5$};           
\end{scope}
\begin{scope}
  \definecolor{strokecol}{rgb}{0.0,0.0,0.0}
  \pgfsetstrokecolor{strokecol}
  \draw (108.0bp,180.0bp) -- (72.0bp,180.0bp) -- (72.0bp,144.0bp) -- (108.0bp,144.0bp) -- cycle;
  \draw (90.0bp,162.0bp) node[font=\LARGE] {1};
\end{scope}
\begin{scope}
  \definecolor{strokecol}{rgb}{0.0,0.0,0.0}
  \pgfsetstrokecolor{strokecol}
  \draw (45.0bp,36.0bp) -- (9.0bp,36.0bp) -- (9.0bp,0.0bp) -- (45.0bp,0.0bp) -- cycle;
  \draw (27.0bp,18.0bp) node[font=\LARGE] {2};
\end{scope}
\end{tikzpicture}}}
  \hspace*{0.4cm}
  \raisebox{4.3cm}{(b)}
  {\normalfont \scalebox{0.4}{\begindg
  \pgfsetlinewidth{1bp}
\pgfsetcolor{black}
  \draw [->] (21.084bp,144.2bp) .. controls (20.28bp,136.18bp) and (20.057bp,126.52bp)  .. (21.105bp,107.59bp);
  \draw [->] (32.916bp,144.2bp) .. controls (33.72bp,136.18bp) and (33.943bp,126.52bp)  .. (32.895bp,107.59bp);
  \draw [->] (21.084bp,72.202bp) .. controls (20.3bp,64.386bp) and (20.069bp,55.017bp)  .. (21.055bp,36.093bp);
  \draw [->] (32.916bp,72.202bp) .. controls (33.7bp,64.386bp) and (33.931bp,55.017bp)  .. (32.945bp,36.093bp);
  \draw [->] (21.084bp,216.2bp) .. controls (20.28bp,208.18bp) and (20.057bp,198.52bp)  .. (21.105bp,179.59bp);
  \draw [->] (32.916bp,216.2bp) .. controls (33.72bp,208.18bp) and (33.943bp,198.52bp)  .. (32.895bp,179.59bp);
  \draw [->] (65.746bp,288.57bp) .. controls (69.398bp,280.35bp) and (73.84bp,270.36bp)  .. (81.991bp,252.02bp);
  \draw [->] (50.496bp,288.57bp) .. controls (46.905bp,280.23bp) and (42.526bp,270.06bp)  .. (34.543bp,251.52bp);
\begin{scope}
  \definecolor{strokecol}{rgb}{0.0,0.0,0.0}
  \pgfsetstrokecolor{strokecol}
  \draw (27.0bp,234.0bp) ellipse (27.0bp and 18.0bp);
  \draw (27.0bp,234.0bp) node[font=\LARGE] {\rnode{$\f{r}_1$}};
\end{scope}
\begin{scope}
  \definecolor{strokecol}{rgb}{0.0,0.0,0.0}
  \pgfsetstrokecolor{strokecol}
  \draw (27.0bp,162.0bp) ellipse (27.0bp and 18.0bp);
  \draw (27.0bp,162.0bp) node[font=\LARGE] {\rnode{$\f{r}_1$}};
  \draw (27.0bp-35bp,162.0bp+15bp) node[font=\Large] {$4$};               
\end{scope}
\begin{scope}
  \definecolor{strokecol}{rgb}{0.0,0.0,0.0}
  \pgfsetstrokecolor{strokecol}
  \draw (27.0bp,90.0bp) ellipse (27.0bp and 18.0bp);
  \draw (27.0bp,90.0bp) node[font=\LARGE] {\rnode{$\f{r}_1$}};
  \draw (27.0bp-35bp,90.0bp+15bp) node[font=\Large] {$3$};               
\end{scope}
\begin{scope}
  \definecolor{strokecol}{rgb}{0.0,0.0,0.0}
  \pgfsetstrokecolor{strokecol}
  \draw (58.0bp,306.0bp) ellipse (27.0bp and 18.0bp);
  \draw (58.0bp,306.0bp) node[font=\LARGE] {\rnode{$\f{r}_0$}};
  \draw (58.0bp-35bp,306.0bp+15bp) node[font=\Large] {$5$};                   
\end{scope}
\begin{scope}
  \definecolor{strokecol}{rgb}{0.0,0.0,0.0}
  \pgfsetstrokecolor{strokecol}
  \draw (108.0bp,252.0bp) -- (72.0bp,252.0bp) -- (72.0bp,216.0bp) -- (108.0bp,216.0bp) -- cycle;
  \draw (90.0bp,234.0bp) node[font=\LARGE] {1};
\end{scope}
\begin{scope}
  \definecolor{strokecol}{rgb}{0.0,0.0,0.0}
  \pgfsetstrokecolor{strokecol}
  \draw (45.0bp,36.0bp) -- (9.0bp,36.0bp) -- (9.0bp,0.0bp) -- (45.0bp,0.0bp) -- cycle;
  \draw (27.0bp,18.0bp) node[font=\LARGE] {2};
\end{scope}
\end{tikzpicture}}}
\end{wrapfigure}

By rewriting schema occurrences in the PS-terms of~(\ref{eq-f8r-ab}) to their
defining CL-term, i.e., by rules $r_0(p,q) \rimp p q$ and $r_1(p,q) \rimp
\c{B} p q$, we obtain (\ref{eq-f8-b-proofs}).(a) for~(\ref{eq-f8r-ab}).(a)
and~(\ref{eq-f8-b-proofs}).(f) for (\ref{eq-f8r-ab}).(b). In general,
rewriting schema occurrences with their defining CL-terms can be applied
individually on each factor, increasing the compacted size only linearly (if
duplicate occurrences of schema parameters are not allowed in defining
CL-terms) and resulting in a CL-term without heuristically motivated proof
schemas, but only combinators in addition to constants representing proper
axioms.

In the proofs~(\ref{eq-f8r-ab}), the schemas occur only instantiated in
specific ways that respect an associated arity, which, in the resolution-like
view is the number of antecedents of the involved Horn clauses. For example,
for $\f{r}_0(p,q)$, the $p$ argument is a term with $\f{r}_1$ as outermost
function symbol and the $q$ argument is either
axiom~$1$ or a term with~$\f{r}_0$ as outermost function symbol. To make use
of this restriction in proof search, the specification of schemas can be
supplemented with \defname{arity types} of parameters and values. Also with
the axioms and the goal such an arity type is then associated, as shown in
Table~\ref{tab-f8-schema-patterns-atypes} for our running example, generalized
to goals $\P(\ff^n(\fa))$ for $n \geq 0$. For these problems, the specified
arity typing has no effect on the found proofs, but lessens the search effort.
In Sect.~\ref{sec-exp-search} we will see a second purpose of arity types, to
disambiguate between object- and meta-level for condensed detachment problems
where implicational formulas are reified.

\begin{table}[h]
  \vspace{-5pt}
  \centering
  \caption{Axiom, schema and goal specifications supplemented by arity
    types.}
  \label{tab-f8-schema-patterns-atypes}
  $\begin{array}{cl}
    \text{\it Axiom ID} & \text{\it Axiom formula}\\\midrule
    1\at 0 & \P(\fa)\\
    2\at 1 & \P(x) \imp \P(\ff(x))
  \end{array}
  \hspace{3.5em}
\begin{array}{llll}
  \text{\it Schema} & \text{\it Defining CL-term}\\\midrule
  \f{r}_0(p\at 1,q\at 0)\at 0 & pq\\
  \f{r}_1(p\at 1,q\at 1)\at 1 & \c{B}pq
\end{array}
  \hspace{3.5em}
  \begin{array}{l}
  \text{\it Goal}\\\midrule
  \P(\ff^n(\fa))\at 0\\
  \ 
  \end{array}$
\end{table}

\enlargethispage{16pt}
\subsection{Practical Realization: The \CCS Prover as a Component of \CDTOOLS}

The approach is realized as a component called \CCS (for \name{Compressed
  Combinatory Structures}) of \CDTOOLS \cite{cw:cdtools}, a Prolog environment
for experimenting with first-order ATP where emphasis is on proof structures
like D-terms as data objects, which in turn is embedded in \PIE
\cite{cw:pie:2016,cw:pie:2020}. \CDTOOLS and \PIE are free software and run in
\name{SWI-Prolog} \cite{swiprolog}. The starting point of \CDTOOLS is
condensed detachment as a specialization of the connection method
\cite{cwwb:lukas:2021}, providing a simplified variant of first-order ATP that
still has many of its essential characteristics and is suitable as basis for
the development and study of new techniques. Condensed detachment has
dedicated applications
\cite{ulrich:legacy:2001,veroff:cd:2011,fitelson:walsh:2021}, reflected in
about 200 \TPTP problems which we so far used in experiments, and can more
generally be used as inference rule for first-order Horn problems.
\CCS currently supports enumeration of proof structures of or up to a given
compacted size, i.e., DAG enumeration, complementing the \SGCD
\cite{cw:cdtools} prover of \CDTOOLS, which enumerates by tree-oriented
measures. Of course, unification is woven into the enumeration, such that only
proofs of a given goal or, if the goal is unspecified, proofs of lemmas
derivable from the axioms are enumerated.

The DAG enumeration method of \CCS keeps a list of proofs of solved subgoals
that are subproofs of the proof under construction. The list is subject to
backtracking. A subgoal is solved by first trying the proofs in the list,
which corresponds in case of success to attaching a new incoming edge to the
DAG node representing the subproof. Then proofs of the subgoal are newly
computed and discarded if they already appear in the list. This is slightly
different from the \name{value-number method}
\cite{aho:compilers:86,genitrini:2020} for DAG construction, where there is no
first step of trying to find a tree in the list and a newly computed tree that
is equal to a tree in the list is not discarded but identified with the tree
in the list (gets the same ``number''). For us, that method seems not
suitable, because adequate size restrictions for the newly computed trees are
then not available when they are created.
 
The basic user input to \CCS is a problem specification along with a
specification of the elements to be used for proof structure construction. The
problem specification determines a set of Horn clauses as proper axioms and
(optionally) a goal atom.\footnote{It possible to run \CCS without given goal
  to generate lemmas.} It can be provided in various formats, e.g., as
identifier of a \TPTP problem, as file in \TPTP format, or as implication in
\PIE's formula syntax. Some formats are subjected to preprocessing, e.g.,
normal form transformation or conversions to handle a non-atomic goal. For
problems that are given as condensed detachment problems, in contrast to
general Horn problems, a special translation is available.
The elements for proof structure construction are in the simplest ``vanilla''
case just the binary $\D$ function symbol and constants for the proper axioms,
sufficient to find pure D-term proofs with minimal compacted size. To make use
of compression, combinators and proof schemas can be specified, where a proof
schema is characterized by a pattern, optionally with arity types for its
variables and result, together with a CL-term or $\lambda$-term that specifies
its semantics.

To keep the interplay between proper axioms, specification of proof term
constructors and handling of lemmas with subproofs for DAG construction
manageable, flexibly configurable and efficient, \CCS is realized as a
compiler that, like \name{PTTP} \cite{pttp:prolog}, generates Prolog code.
\CDTOOLS further includes supplementary functionality for experimenting with
the approach, including graph-based normal form conversion for CL-terms and an
interface to \name{TreeRePair} \cite{lohrey:treerepair:2013}, an external tool
for advanced tree compression.

\section{First Experiments}
\label{sec-experiments}

The following three subsections each describe a series of experiments based on
the corpus \TPTPCDT, which comprises the 196 condensed detachment problems in
\name{TPTP~7.5.0} that remain after excluding from all 206~condensed
detachment problems those two with status \name{satisfiable}, those five with
a form of detachment that is based on implication represented by disjunction
and negation, and those three with a non-atomic goal theorem.\footnote{The
  restriction on detachment and goals simplifies the problem form and was used
  in previous experiments \cite{cw:cdtools}. The eight problems that do not
  match it are suitable as inputs of \CCS in a mode for general Horn
  problems.}
References to the \name{rating} of a problem refer to the latest rating in
\name{TPTP~7.5.0}. While some of the experiments include proofs for about 25
problems rated between 0.25 and 0.50, thus comparing to results of
state-of-the art solvers, others explore what can be reached by exhaustive
search upon compacted size, which seems useful, e.g., to find guaranteed
shortest proofs, to gain an overview on redundancies in the search space, and
as a basis for refinements. For the latter type of experiments the range is
around 80--90 proven problems, which actually is at the top of what is known
for clausal tableau provers.\footnote{\name{leanCoP~2.1} \cite{leancop} proves
  50 problems in the setting of the experiments reported here. The
  \name{ProblemAndSolutionStatistics} document of \name{TPTP 8.0.0} reports
  the following numbers of proved problems: \name{SETHEO~3.3}
  \cite{setheo:3.3:1997}: 65; \name{S-SETHEO} \cite{s-setheo:2001}: 66;
  \name{lazyCoP~0.1} \cite{lazycop:2021}: 42; \name{SATCoP~0.1}
  \cite{satcop:2021}: 59; all four together: 76. \CMProver \cite{cw:pie:2016}
  in various configurations proves 89~problems
  (\url{http://cs.christophwernhard.com/pie/cmprover/evaluation_201803/tptp_neq.html}).
  All these systems together prove~92 problems.}

All results were obtained on a HPC system with
\name{Intel\textregistered\ Xeon\textregistered\ Platinum 9242 @ 2.30GHz}
CPUs, 3.7~GB memory per CPU and 2,400~s time limit per prover run and problem.
In their current versions the used provers \CCS and \SGCD do not support
parallelism or portfolio configurations. If adequate, we consider results of
several configurations joined together, picking the best result for each
problem, as if obtained in a parallel run. For such combined configurations,
each individual configuration was given the 2,400~s time limit.

The \CDTOOLS web page shows detailed tables with links to the individual \TPTP
problems and to graphical proof structure presentations. It also archives
downloadable logs that contain the proofs from the experiments in
Prolog-readable form.

\subsection{Pure D-Term Proofs with Minimal Compacted Size}
\label{sec-exp-min-csize}

\CCS was applied in ``vanilla'' configurations to determine by exhaustive
search as a far as possible for each problem in the corpus the minimal
compacted size of a pure D-term proof and, moreover, the number of different
such proofs of that size. The results give an overview on the range of such
exhaustive search and provide guiding values for comparison with compressed
proofs. Table~\ref{tab-mincsize} aggregates the results. \CCS was used in two
configurations that differed in the order in which proper axioms and $\D$ are
considered at enumeration. The table shows in each row for a set of problems
its cardinality and, aggregated as minimum, maximum, average and median over
the set, the number of proper axioms (not counting the detachment clause and
the negated goal clause), the minimal compacted size of a pure D-term proof,
and the number of different pure D-term proofs with that size. The problem
sets are as follows: \TPTPCDT is the corpus of considered problems, \name{MC}
is the set of problems in the corpus for which \CCS (in at least one of the
two configurations) found the minimal compacted size, \name{MC all} is the
subset of \name{MC} for which \CCS could determine the number of all proofs of
that size, and \name{Not MC} is \TPTPCDT without the members of \name{MC}. For
these just lower size bounds could be determined. Three problems in \name{MC}
and two in \name{MC all} are rated~0.25, the remaining ones in \name{MC} are
rated~0.00.

\begin{table}[h]
  \centering
  \caption{Determined minimal compacted size and number of different proofs of
    that size for \TPTPCDT.}
  \label{tab-mincsize}
  \setlength{\tabcolsep}{3pt}
\begin{tabular}{lr@{\hspace{15pt}}rrrr@{\hspace{15pt}}rrrr@{\hspace{15pt}}rrrr}
  \multirow{2}{*}{\it Problem set}  & \multirow{2}{*}{\it \#} &
  \multicolumn{4}{c}{\it \#Axioms} &
  \multicolumn{4}{c}{\it Min. comp. size} & \multicolumn{4}{c}{\it \#Proofs of min. comp. size}\\
                            &   & min & max & avg & med & min & max & avg &  med & min & max & avg & med\\\midrule
  \TPTPCDT                  &  196 & 1 & 5 & 2.46 & 3\\
  \name{MC}                        &   86 & 1 & 5 & 2.33 & 2 &  1 & 12 & 6.41 & 7\\
  \name{MC all}                      &   79 & 1 & 5 & 2.30 & 2 &  1 & 12 & 6.09 & 6   & 1 & 17,280 & 368.08 & 3\\
  \name{Not MC}                     &  110 & 1 & 4 & 2.56 & 3 &  $\geq$8 & $\geq$12 & $\geq$9.31 & $\geq$9
\end{tabular}
\end{table}

\subsection{Combinatory Compression of Given Proofs}
\label{sec-exp-compress}

In this experiment we first converted given proofs of problems from the corpus
to compressed tree grammars, with an advanced tool originally targeted at
compressing XML trees, and then translated the grammars into CL-terms, via
$\lambda$-terms and an optimized technique from the implementation of
functional programming languages. Specifically, the grammar compression was
obtained with \name{TreeRePair} \cite{lohrey:treerepair:2013} with option
\name{-optimize edges}. It compresses the given binary tree representation of
the proof into an acyclic context-free grammar with exactly one production for
each nonterminal. The nonterminals may have parameters, which have only a
single occurrence in the right-hand side (i.e., the grammars are
\name{linear}). We translated the productions first into $\lambda$-terms and
then into CL-terms by the optimized method from
\cite[Chap.~16]{peytonjones:87}, followed by some graph-based simplification.
Table~\ref{tab-combinators} shows all combinators considered there.
\begin{table}[h]
  \centering
\caption{Considered combinators with defining $\lambda$-term and axiom formula
  (principal type). For some, an alternate definition in terms of other
  combinators is given. The 4-ary combinators $\c{S_4}$, $\c{B_4}$, $\c{C_4}$
  appear as $\c{S'}$, $\c{B^*}$,$\c{C'}$ in \cite{peytonjones:87}, which
  clashes with other uses of these names in the literature. $\c{B}\c{B}$ has
  no widespread name.}
\label{tab-combinators}
$\begin{array}{llll}
        & \text{$\lambda$-Term} & \text{Principal Type} &
  \text{Alt.} \\\midrule
  \c{S} & \lambda xyz \ldot x z (y z)
  & (p\oimp (q\oimp r))\oimp ((p\oimp q)\oimp (p\oimp r))\\
  \c{K} & \lambda xy \ldot x
  & p\oimp (q\oimp p)\\
  \c{I} & \lambda x \ldot x
  & p\oimp p\\  
  \c{B} & \lambda xyz \ldot x (y z)
  & (p\oimp q)\oimp ((r\oimp p)\oimp (r\oimp q))\\
  \c{C} & \lambda xyz \ldot x z y
  & (p\oimp (q\oimp r))\oimp (q\oimp (p\oimp r))\\
  \c{S_4} &  \lambda xyzu . x (y u) (z u)
  & (p\oimp (q\oimp r))\oimp ((s\oimp p)\oimp ((s\oimp q)\oimp (s\oimp r)))\\
  \c{B_4} &  \lambda xyzu . x (y (z u))
  & (p\oimp q)\oimp ((r\oimp p)\oimp ((s\oimp r)\oimp (s\oimp q)))\\
  \c{C_4} & \lambda xyzu .  x (y u) z
  & (p\oimp (q\oimp r))\oimp ((s\oimp p)\oimp (q\oimp (s\oimp r)))\\
  \c{I'} & \lambda xy \ldot y x
  & p\oimp ((p\oimp q)\oimp q)
  & \c{C} \c{I}\\
  \c{B'} & \lambda xyz \ldot y (x z)
  & (p\oimp q)\oimp ((q\oimp r)\oimp (p\oimp r)))
  & \c{C} \c{B}\\
  \c{C^*} & \lambda xyz \ldot y z x
  & p\oimp ((q\oimp (p\oimp r))\oimp (q\oimp r)))
  & \c{C} \c{C}\\
  \c{B^{''}} & \lambda xyzu \ldot x y (z u)
  & (p\oimp (q\oimp r))\oimp (p\oimp ((s\oimp q)\oimp (s\oimp r)))
  & \c{B} \c{B}
\end{array}$
\end{table}

Our objectives with this experiment were to get an idea about what can be
achieved for proofs stemming from applications with compression techniques and
to see which combinators emerge, as potential guidance for selecting
combinators in proof search. We were interested in the ``real'' potential of
compression, rather than just effects on redundancies that are trivial or
better to detect with other means. Hence the given proofs should already be
small. For this purpose, they were obtained by \SGCD \cite{cw:cdtools} and
\CCS. Proofs for 167 problems were found in settings where the size of the
CL-term translation is recorded and only a single proof with the smallest
CL-translation that was found before the timeout is kept. These proofs were
supplemented by further proofs from previous experiments with \SGCD and \CCS
to form a basis of 217~proofs for 176~problems of the corpus, including 25
rated larger than~0.00, with 5~rated~0.50, but none~1.00.
Table~\ref{tab-compress-combs} shows the combinators, or, more precisely,
\emph{maximal pure CL-terms}, i.e., pure CL-terms in occurrences not as strict
subterm of another pure CL-term, in the CL-translations obtained in the
experiment. Note that $\c{S}$ and $\c{S_4}$, whose defining $\lambda$-term has
a variable with two occurrences in its body, do not appear in the proofs,
because the underlying grammars were linear.
\begin{table}[h]
  \centering
  \caption{Maximal pure CL-terms occurring in the CL compressed proofs, with
    the number of proofs in which they occur (among 132 proofs for 86 problems
    in which combinators occur in the CL compression).}
  \label{tab-compress-combs}
  \setlength{\tabcolsep}{3pt}
\begin{tabular}{lr@{\hspace{3em}}lr@{\hspace{3.2em}}lr@{\hspace{3.2em}}lr@{\hspace{3.2em}}lr@{\hspace{3.2em}}lr} 
  $\c{I^{\prime}}$ & 42 & $\c{C}$    & 28 & $\c{C_{4}}$ & 5
  &  $\c{C^{*}}$ & 2 & $\c{B_{4}}\c{B_{4}}\c{B}$ & 2 & $\c{C_{4}}\c{C}$ & 1\\
  $\c{B}$        & 37 & $\c{B_{4}}$ & 9  & $\c{B^{\prime}}$  & 2
  & $\c{B_{4}}\c{B}\c{B}$ & 2 & $\c{C_{4}}\c{B}$ & 1 & $\c{C_{4}}\c{B}\c{B}$ & 1
\end{tabular}
\vspace{-10pt}
\end{table}

Tables~\ref{tab-compress-summary} and~\ref{tab-compress-ratios} summarize the
achieved compression effects, where we use the following shorthands for size
measures of a proof in its different transformations: the originally given
pure D-term, the compressed tree grammar and the compressed CL-term.
\begin{itemize}
\item[\LC:] Compacted size of the compressed CL-term, where maximal pure
  CL-terms are valued like constants, reflecting that a pure CL-term may be
  expressed by a dedicated combinator.
\item[\XC:] Compacted size of the given pure D-term.
\item[\GS:] Size of the compressed tree grammar, determined as the sum over
  the number of edges in all its right-hand sides. (If the grammar has only
  parameter-free nonterminals, it represents a DAG. \GS\ is then the doubled
  compacted size of the represented term.)
\end{itemize}

\begin{table}[h]
  \vspace{-12pt}
  \centering
  \caption{Problems with given proofs on which compression to CL-terms and
    tree grammars had reducing effect.}
  \label{tab-compress-summary}
  \begin{tabular}{lrrrrrr}
    {\it Problem set}            & {\it \# } & {\it \LC$<$\XC}  &
    {\it CL-term has comb.} &
    {\it \GS$<$\XC$*$2} & {\it Gr. has param.}\\\midrule
    \TPTPCDT                   & 196 &         &         &\\
    \name{Proven}              & 176 &      29 &  86 &       110 &         133\\
    \name{MC}                  &  87 &       1 &  16 &        32 &          45\\
    \TPTPCDT, \name{Rtg$>$0.00} &  45 &         &     &           &            \\
    \name{Proven, Rtg$>$0.00}   &  25 &      11 &  21 &        22 &          24  
  \end{tabular}
\end{table}
\begin{table}[h]
  \vspace{-15pt}
  \centering
  \caption{Compression ratios for problems where the compression is strictly
    smaller than the source.}
  \label{tab-compress-ratios}
  \begin{tabular}{lrrrrr}
    {\it Problem set} & {\it \# } & min & max & avg & med\\\midrule
    \name{Proven, \LC$<$\XC}              &  29 & 1.03 & 1.17 & 1.09 & 1.10\\
    \name{Proven, \LC$<$\XC, Rtg$>$0.00}   &  11 & 1.03 & 1.12 & 1.08 & 1.07\\
    \name{Proven, \GS$<$\XC$*$2}            & 110 & 1.02 & 1.83 & 1.26 & 1.21\\
    \name{Proven, \GS$<$\XC$*$2, Rtg$>$0.00} &  22 & 1.02 & 1.56 & 1.32 & 1.38
  \end{tabular}
  \vspace{-3pt}
\end{table}

\enlargethispage{10pt}

Table~\ref{tab-compress-summary} shows in each row for a set of problems its
cardinality, the number of members with \LC$<$\XC, the number of members where
a combinator occurs in the compressed CL-term, the number of members with
\GS$<$\XC$*$2, i.e., where the grammar compression is smaller than the DAG
considered as grammar, and the number of members where the grammar has a
parameterized nonterminal, indicating that the grammar compression did not
effect just DAG compression. (The number of parameters of nonterminals in all
obtained grammars is between~0 and~3.) The problem sets are as follows:
\TPTPCDT is the corpus of considered problems. \name{Proven} is the set of
problems for which a proof was available. \name{MC} is the set of problems for
which the minimal compacted size of a pure D-term proof is
known.\footnote{Compared to the set \name{MC} of Table~\ref{tab-mincsize} it
  contains one more problem, where the techniques considered there could
  ascertain only a lower bound of the compacted size and a proof of that size
  was found with another technique.} \TPTPCDT \name{Rtg$>$0.00} and
\name{Proven} \name{Rtg$>$0.00} are like \TPTPCDT and \name{Proven}, but
restricted to problems rated larger than~0.00.
Apparently compressions have on higher rated problems a stronger effect, in
particular compared to those where the minimal compacted size of pure D-terms
can be determined.
A size reduction achieved by the grammar compression (e.g., for 110 of the 176
problems in \name{Proven}) is not always reflected in the introduction of a
combinator (only for 86 problems in \name{Proven}) and a size reduction of the
CL-term through combinators (only for 29 of these 86 problems). This needs
further investigation. It may be due to a linear size increase in the grammar
to combinator translation and the way combinators themselves are counted.
Table~\ref{tab-compress-ratios} shows in each row for a set of problems its
cardinality and compression ratios, aggregated as minimum, maximum, average
and median over the set.
The problem sets are as follows: \name{Proven} is the set of problems for
which a proof is available and it holds that {\it \LC$<$\XC}, i.e., the
combinatory compression has reducing effect. The considered compression ratio
for this set is {it \XC/\LC}. \name{Proven, \GS$<$\XC$*$2} is the set of
problems for which a proof is available and it holds that \GS$<$\XC$*$2,
i.e., the grammar compression taken as basis for the combinatory compression
has reducing effect. The considered compression ratio is {\it \XC$*2/$\GS}.
\name{Proven, \LC$<$\XC, Rtg$>$0.00} and \name{Proven, \GS$<$\XC$*$2,
  Rtg$>$0.00} are like the other sets, but restricted to problems rated larger
than~$0.00$. The modest compression ratios do not come as a surprise since the
problems are not constructed challenges but stem from applications that where
previously addressed by ATP systems. As discussed in the context of
Table~\ref{tab-compress-summary}, the reducing effects of the grammar
compression are not in full reflected in the combinatory translation, which
needs further investigation.

\subsection{Exhaustive Search with Proof Schemas Involving Combinators}
\label{sec-exp-search}

\enlargethispage{10pt}

\CCS was applied with combinator-based proof schemas and exhaustive
enumeration of PS-terms by iterative deepening upon compacted size. The main
objective was to see the basic effects of proof search with compressed
combinatory structures for problems from applications. Does the implicitly
incorporated search for shortest compressions obstruct proof search or does it
increase success through the shorter structures? The used configurations of
\CCS were characterized by subsets of the schemas from
Table~\ref{tab-proofschemas-search}. Arity type~$0$ is used for proper axioms
and the goal. $\c{I}$ with arity type~$2$ together with $\f{r}_1$ expresses
detachment: A schema instance $\f{r}_1(\c{I}, p, q)$ expands into $\c{I} p q$,
which reduces into $p q$, or $\D(p,q)$. Similarly, detachment can be expressed
with $\c{I}$, $\f{r}_0$ and $\f{r}_2$.
\begin{table}[h]
  \vspace{-5pt}
  \centering
  \caption{Considered proof schemas.}
  \label{tab-proofschemas-search}
  \setlength{\arraycolsep}{3.5pt}
$\begin{array}{llll}
  \text{Schema} & \text{CL-term} & \text{$\lambda$-term} & 
  \text{Resolution-like view}\\\midrule
  \c{I}\at 2 & \c{I} & \lambda x y . x y &
  \derives (A \nimp B) \nimp (A \nimp B)\\
  \f{r}_0(p\at 1,q\at 0)\at 0 & pq
  & pq
  & A \nimp B,\, D\; \derives\; B\mgu(\{A, D\}) \\
  \f{r}_1(p\at 2,q\at 0,r\at 0)\at 0 & p q r &  p q r &
  A_1 \nimp (A_2 \nimp B),\, D_1,\, D_2\; \derives\; B\mgu(\{\{A_1,D_1\},\{A_2,D_2\}\})\!\!\\
  \f{r}_2(p\at 2,q\at 0)\at 1 & pq
  & \lambda x . pqx
  & A_1 \nimp (A_2 \nimp B),\, D\; \derives\; (A_2 \nimp B)\mgu(\{A_1, D\}) \\
  \f{r}_3(p\at 2,q\at 0)\at 1 & \c{C}pq
  & \lambda x . pxq
  & A_1 \nimp (A_2 \nimp B),\, D\; \derives\; (A_1 \nimp B)\mgu(\{A_2, D\}) \\
  \f{r}_4(p\at 1, q\at 1)\at 1 & \c{B}pq
  & \lambda x. p(qx)
  & A \nimp B,\, C \nimp D\; \derives\; (C \nimp B)\mgu(\{A, D\}) \\
  \f{r}_5(p\at 2, q\at 1)\at 2 & \c{B}pq
  & \lambda xy. p(qx)y
  & A_1 \nimp (A_2 \nimp B),\, C \nimp D\; \derives\; (C \nimp (A_2 \nimp B))\mgu(\{A_1,D\}) \\
  \f{r}_6(p\at 2, q\at 1)\at 2 & \c{B} (\c{C} p) q
  & \lambda xy . py(qx)
  & A_1 \nimp (A_2 \nimp B),\, C \nimp D\; \derives\; (C \nimp (A_1 \nimp B))\mgu(\{A_2,D\}) \\
\end{array}$
\end{table}

The enumerated proof structures are PS-terms built-up from constants for
proper axioms and the configured schemas. Finally, they were subjected to a
simplification where schemas whose definition involves no combinator are
rewritten into their definiens, which often leads to a slight size
reduction.\footnote{With the rules $\f{r}_0(p, q) \rimp p q$, $\f{r}_2(p, q)
  \rimp p q$, $\f{r}_1(i, p, q) \rimp p q$, $\c{I} p \rimp p$. This induces a
  slight mismatch between the assessed compacted size during enumeration and
  that of the final proof.} The final PS-term then may also contain
occurrences of application as binary symbol.
Tables~\ref{tab-search-solutions}-- \ref{tab-search-smx} summarize the
outcomes, where we use the following shorthands for size measures of a given
proof expressed as PS-term.
\begin{itemize}
\item[\SC:] Compacted size of the PS-term, i.e., the number of inner nodes
  of its minimal DAG.
\item[\XC:] Compacted size of the pure D-term obtained from the PS-term by
  expanding  schemas and rewriting combinators.
\item[\MC:] Minimal compacted size of a pure D-term that proves the same
  problem as the given proof, if known.
\end{itemize}

Table \ref{tab-search-solutions} shows in each row for a configuration of \CCS
the number of problems that could be proven, the number of problems with a
proof in which $\f{r}_3$, $\f{r}_4$, $\f{r}_5$, or $\f{r}_6$ (i.e., a schema
involving a combinator other than $\c{I}$) occurs, and two compression ratios,
aggregated as minimum, maximum, average and median. The first is the ratio
\XC/\SC\ for the proofs involving $\f{r}_3$, $\f{r}_4$, $\f{r}_5$, or
$\f{r}_6$. Then the number of problems in this set for which in addition
\MC\ is known is shown, followed by the ratio \MC/\SC\ upon that set.
The considered prover configurations are as follows: \name{S1}, \name{S2}
and~\name{S3} enumerate upon compacted size, building-up PS-terms from the
indicated schemas, $\c{I}$ and identifiers of proper axioms. \name{S1+S2+S3}
combines the best results from \name{S1}, \name{S2} and \name{S3}, ordered by
\SC\ and required run time. \name{\PMC} combines the two ``vanilla''
configurations described in Sect.~\ref{sec-exp-min-csize} and is included for
comparison. All configurations together can prove~92 problems, 89~rated 0.0
and 3~rated 0.25, proved by both, \name{S1+S2+S3} and \name{\PMC}.

\begin{table}[h]
  \centering
  \caption{Number of proven problems, proofs with combinators, and compression
    ratios.}
  \label{tab-search-solutions}
  \setlength{\tabcolsep}{3pt}
  \begin{tabular}{lrr@{\hspace{15pt}}rrrrr@{\hspace{15pt}}rrrr}
    \multirow{2}{*}{\it CCS Configuration} &
    \multirow{2}{*}{\it \#Proven} & \multirow{2}{*}{\it \#Cmb.} &
    \multicolumn{4}{c}{\it \XC/\SC\hspace*{20pt}} & {\it \#Cmb.} &
    \multicolumn{4}{c}{\it \MC/\SC}\\
    &&& {\it min} & {\it max} & {\it avg} & {\it med} & {\it $\cap$\PMC} & {\it min} & {\it max}
    & {\it avg} & {\it med}\\\midrule
    \name{S1+S2+S3} & 88 & 25 & 1.11 & 1.64 & 1.37 & 1.33
    & 19   & 1.11 & 1.40 & 1.28 & 1.29\\
    \name{S1} ($\f{r}_1,\f{r}_2,\f{r}_4,\f{r}_5, \f{r}_6$) & 86 & 20
    & 1.20 & 1.63 & 1.37 & 1.33
    & 16   & 1.20 & 1.40 & 1.30 & 1.31\\
    \name{S2} ($\f{r}_1,\f{r}_2,\f{r}_3,\f{r}_4,\f{r}_5,\f{r}_6$) & 83 & 20
    & 1.11 & 1.57 & 1.31 & 1.33
    & 19   & 1.11 & 1.40 & 1.26 & 1.29\\
    \name{S3} ($\f{r}_0,\f{r}_2,\f{r}_4,\f{r}_6$) & 64 & 30
    & 0.71 & 2.00 & 1.17 & 1.15
    & 29   & 0.67 & 1.17 & 0.96 & 1.00\\
    \name{\PMC} & 86
    \\\midrule
  Total & 92
  \end{tabular}
\end{table}

Of the proof schemas $\fr_3$--$\fr_6$, schema~$\fr_4$ is by \name{S1} and
\name{S2} together used only for 3 problems, but by \name{S3} for 24 problems;
$\fr_3$ is used by \name{S2} for 7 problems and~$\fr_5$ and $\fr_6$ are used
by each of the configurations where they are specified for 10--16 problems.
For~\name{S3} some compression ratios are less than 1.00, indicating that the
compacted size of some PS-terms is larger than that of their pure D-term
normalization or the minimal compacted size for the problem. It remains to
investigate whether this has a substantial reason or can be remedied by some
postprocessing simplification.

\name{S1+S2+S3} failed on four problems proven by \name{\PMC} but proved
six~problems on which \name{\PMC} failed, indicating that for the latter the
search for compressed structures was beneficial. They are shown in
Table~\ref{tab-search-only-s123}, with the number of proper axioms, the \CCS
configuration that provided the proof and parameters of the proof.
Table~\ref{tab-search-smx} shows those problems for which a proof was found by
\name{S1+S2+S3} that satisfies \SC$<$\MC$<$\XC, where \SC\ is minimal among
the proofs by \name{S1+S2+S3}. The relationship \SC$<$\MC$<$\XC\ indicates
that the minimal compacted size of a pure D-term can be undercut by a PS-term
whose pure D-term normalization is larger than the minimal one. All problems
in Table~\ref{tab-search-smx} are hard for conventional tableau
provers.\footnote{\name{ProblemAndSolutionStatistics} of \name{TPTP 8.0.0}
  does contain for neither of the provers \name{SETHEO~3.3}, \name{S-SETHEO},
  \name{lazyCoP~0.1} and \name{SATCoP~0.1} a success record on these problems.
  Also \name{leanCoP~2.1} fails on them with a 2 hour timeout per problem.
  \name{CMProver}, in some configurations, can solve LCL364-1, LCL092-1, and
  LCL365-1.}
\begin{table}[h]
  \vspace{-3pt}
  \centering
  \caption{Proven by S1+S2+S3 but not by \PMC}
  \label{tab-search-only-s123}
  \begin{tabular}{lrrlrrrr}
    {\it Problem} &  {\it Rating}  & {\it \#Ax} & {\it Config} & {\it Time} & \SC & {\it
      \XC} & {\it \XC/\SC}\\\midrule
    LCL080-2 & 0.00 & 4 & \name{S1} & 2,201~s & 7  & 9  & 1.29\\
    LCL088-1 & 0.00 & 1 & \name{S2} & 1,029~s & 10 & 12 & 1.20\\
    LCL090-1 & 0.00 & 1 & \name{S3} & 1,254~s & 11 & 18 & 1.64\\
    LCL366-1 & 0.00 & 3 & \name{S1} & 1,743~s & 8 & 13  & 1.63\\
    LCL378-1 & 0.00 & 3 & \name{S1} & 1,717~s & 8 & 13  & 1.63\\
    LCL399-1 & 0.00 & 3 & \name{S1} & 1,912~s & 8 & 12  & 1.50
  \end{tabular}
\end{table}
\begin{table}[h]
  \vspace{-16pt}  
  \centering
  \caption{Proofs where \SC$<$\MC$<$\XC\ and \SC\ is minimal among the
    considered proofs.}
  \label{tab-search-smx}
  \begin{tabular}{lrrlrrrrrr}
    {\it Problem} & {\it Rating} & {\it \#Ax} & {\it Config} & {\it Time} & \SC & {\it
      \MC} & \XC & {\it \MC/\SC} & {\it \XC/\SC} \\\midrule
    LCL089-1 & 0.00 & 1 & \name{S2} & 85~s & 9 & 10 & 13 & 1.11 & 1.44 \\
    LCL129-1 & 0.00 & 1 & \name{S1} & 48~s & 8 & 11 & 12 & 1.38 & 1.50 \\
    LCL364-1 & 0.00 & 3 & \name{S1} & 70~s & 7 &  9 & 11 & 1.29 & 1.57\\
    LCL092-1 & 0.25 & 1 & \name{S2} & 357~s & 9 & 12 & 13 & 1.33 & 1.44\\
    LCL365-1 & 0.25 & 3 & \name{S1} & 270~s & 8 & 10 & 12 & 1.25 & 1.50
  \end{tabular}
  \vspace{-3pt}
\end{table}

\section{Conclusion}
\label{sec-conclusion}

A new perspective for clausal tableau techniques in first-order ATP has been
shown, where proof search is not performed directly by enumerating tableau
structures, but by enumerating more succinct combinator terms that normalize
into the tableau structures. Restricting combinator terms to configured
patterns provides a proof schema mechanism that constrains proof structures
considered at search. Initial experiments indicate feasibility of the approach
for problems from applications, and special relevance if the objective is to
find particularly short or ``good''\footnote{Aside of measures based directly
  on size of the proof structure
  \cite{veroff:shortest:2001,cwwb:lukas:2021,alrabbaa:goodproofs:2021},
  qualities of interest are, for example, short formula size of lemmas;
  restrictions on the inference rules, e.g., to just condensed detachment
  \cite{veroff:cd:2011}; that extracted interpolants are in some specific
  formula class \cite{benedikt:book,toman:2015:tableaux}; or capturing short
  informal arguments \cite{heule:checkerboard:19}.} proofs, or to get an
overview on different proofs for a problem.

Of course, it would be very interesting to see the approach with problem
series that separate proof systems by their strength. Unfortunately, it seems
that, differently from propositional logic, for first-order Horn formulas such
problems are hardly available. Our running example was from
\cite{eder:relative:1992}, where it separates conventional connection calculi
(including clausal tableaux) on the one hand from resolution and the
connection structure calculus
\cite{eder:cs:1989,eder:relative:1992,bibel:eder:1993} on the other hand. The
presented approach actually emerged from an effort to implement the connection
structure calculus, which was never implemented so far. A precise comparison
of its labeling techniques with our term-oriented view would, however, need
further investigation.

Our proof schemas resemble, at least in part, restricted forms of resolution
for Horn clauses. It is indeed straightforward to express inference rules such
as hyperresolution or binary resolution for clauses of specific lengths by
proof schemas. Determining actual correspondences to known resolution
refinements needs further work. More generally, the potential and expressive
power of combinator terms as compressed and enumerable proof representations
is yet to be explored. Is it possible to capture induction through schemas
based on combinators with repeated variables in the defining $\lambda$-term?
The proof schema mechanism suggests a two-leveled search, some systematic
enumeration of sets of permitted schemas and then the actual proof search with
these. Also enumeration itself can be addressed, with incomplete ways (e.g.,
based on the subproof relationship \cite{cwwb:lukas:2021,cw:cdtools}) and
driven by learned structure patterns.

The problem of finding a proof with minimal compacted size, considered in
Sect.~\ref{sec-exp-min-csize}, is also the subject of
\cite{veroff:shortest:2001}, where a method based on linked UR-resolution and
implemented with \OTTER \cite{otter} is introduced. Apparently, its
performance on \TPTP problems is not known.
Grammar compressions are used in first-order proof theory on the formula level
\cite{hetzl:tree:2012}. Their use for proof structures seems new. Also the
correspondence between grammar compressions and combinator terms seems a new
observation.
Combinator terms are used for \emph{search} in higher-order unification
\cite{dougherty:1993,bhayat:reger:2019} and a related recent superposition
calculus for higher-order logic \cite{bhayat:reger:2021} involves rewrite
rules that realize combinator axioms.

So far, the experiments were performed on the condensed detachment problems in
the \TPTP. As indicated with our running example, the approach extends to
general first-order Horn problems, which has been already laid out in the
implementation. The experiments so far were centered on exhaustive DAG
enumeration. On the agenda is to use the combinator approach with features
previously explored for a related prover \cite{cw:cdtools}, in particular,
with blending goal- and axiom-directed search, tree- instead of DAG-oriented
enumeration, and heuristic restrictions (e.g., on the size of involved lemma
formulas, number of kept lemmas, or by keeping at most a single proof per
lemma). Specifying adequate size measures of proof terms may need further
investigation: It appears that details which from a complexity point of view
play no role show up as confusing when single proofs are considered. Possibly
this is a matter of graph-based postprocessing simplifications. From a broader
perspective, the approach opens tempting bridges between first-order ATP and
neighboring fields, such as type theory
\cite{hindley:meredith:cd:1990,hindley:book:1997} and the recent empirical
computation theory based on enumerating combinator terms \cite{wolfram:2021}.

\enlargethispage{5pt}
\begin{acknowledgments}
  The author thanks Wolfgang Bibel for inspirations and for sharing an idea --
  more directly based on the connection method -- to prove a problem series,
  which gave impetus to this work. The author also thanks anonymous reviewers
  for helpful feedback to improve the presentation. Funded by the Deutsche
  Forschungsgemeinschaft (DFG, German Research Foundation) --
  Project-ID~457292495. The work was supported by the North-German
  Supercomputing Alliance (HLRN).
\end{acknowledgments}

\bibliography{biblukas03ed}

\end{document}